\documentclass[fleqn,10pt]{wlscirep}

\title{Element-Specific Orbital Character in a Nearly-Free-Electron Superconductor Ag$_5$Pb$_2$O$_6$ Revealed by Core-Level Photoemission}

\author[1,2]{Soobin Sinn}
\author[3]{Kyung Dong Lee}
\author[3]{Choong Jae Won}
\author[1,2]{Ji Seop Oh}
\author[4]{Moonsup Han}
\author[4]{Young Jun Chang}
\author[3]{Namjung Hur}
\author[5]{Byeong-Gyu Park}
\author[1,2]{Changyoung Kim}
\author[1,2,*]{Hyeong-Do Kim}
\author[1,2,**]{Tae Won Noh}

\affil[1]{Center for Correlated Electrons Systems, Institute for Basic Science (IBS), Seoul 08826, Republic of Korea}
\affil[2]{Department of Physics and Astronomy, Seoul National University (SNU), Seoul 08826, Republic of Korea}
\affil[3]{Department of Physics, Inha University, Incheon 22212, Republic of Korea}
\affil[4]{Department of Physics, University of Seoul, Seoul 02504, Republic of Korea}
\affil[5]{Pohang Accelerator Laboratory, Pohang University of Science and Technology, Pohang 37673, Republic of Korea}

\affil[*]{hdkim6612@snu.ac.kr}
\affil[**]{twnoh@snu.ac.kr}

\begin{abstract}
Ag$_5$Pb$_2$O$_6$ has attracted considerable attention due to its novel nearly-free-electron superconductivity, but its electronic structure and orbital character of the Cooper-pair electrons remain controversial.
Here, we present a method utilizing core-level photoemission to show that Pb 6$s$ electrons dominate near the Fermi level.
We observe a strongly asymmetric Pb 4$f_{7/2}$ core-level spectrum, while a Ag 3$d_{5/2}$ spectrum is well explained by two symmetric peaks.
The asymmetry in the Pb 4$f_{7/2}$ spectrum originates from the local attractive interaction between conducting Pb 6$s$ electrons and a Pb 4$f_{7/2}$ core hole, which implies a dominant Pb 6$s$ contribution to the metallic conduction.
In addition, the observed Pb 4$f_{7/2}$ spectrum is not explained by the well-known Doniach-\v{S}unji\'{c} lineshape for a simple metal.
The spectrum is successfully generated by employing a Pb 6$s$ partial density of states from local density approximation calculations, thus confirming the Pb 6$s$ dominant character and free-electron-like density of states of Ag$_5$Pb$_2$O$_6$.
\end{abstract}

\begin{document}

\flushbottom
\maketitle

\thispagestyle{empty}

\section*{Introduction}

\begin{figure}
\centering
\includegraphics{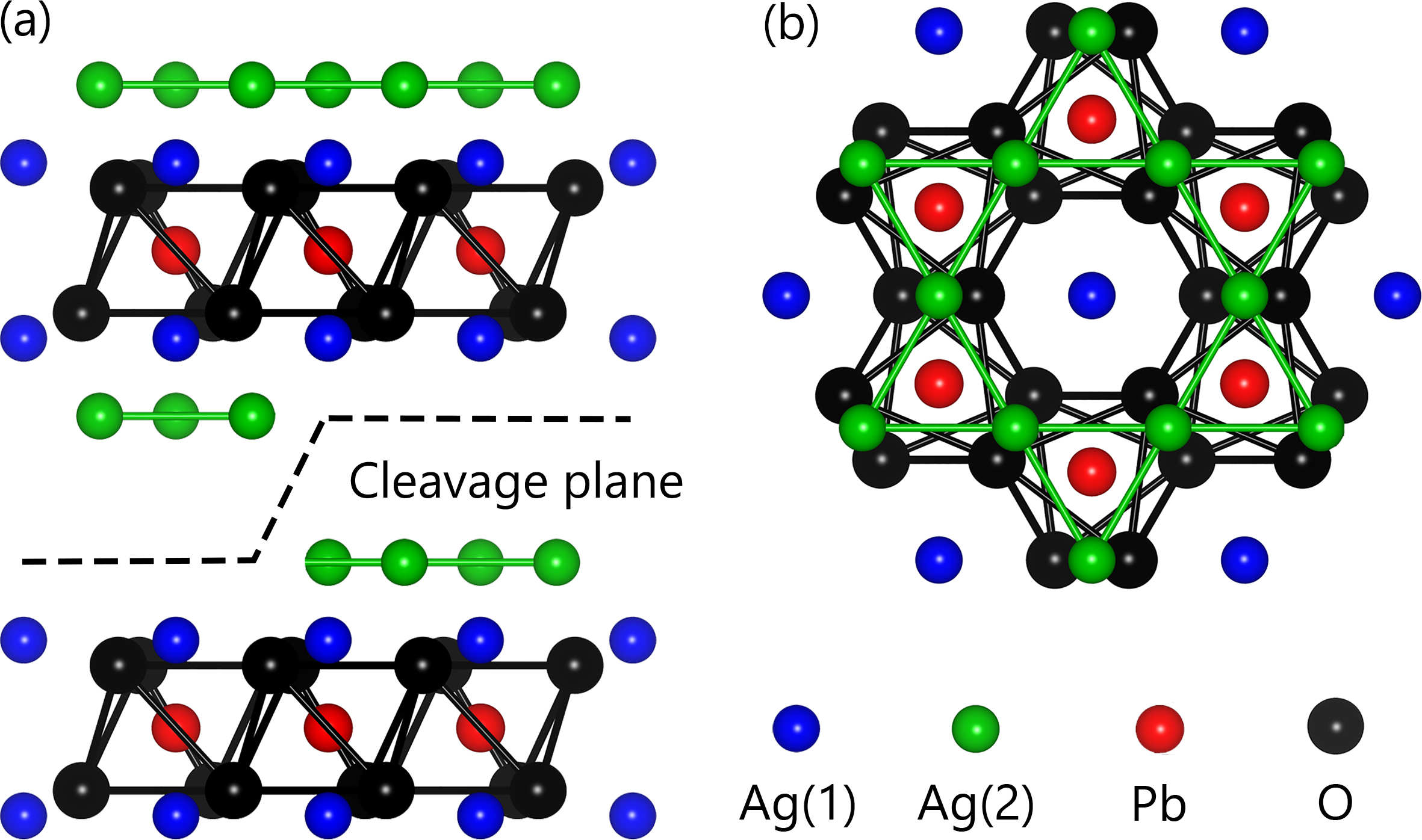}
\caption
{
(a) Side view and (b) top view of the crystal structure of Ag$_5$Pb$_2$O$_6$.
Green, blue, red, and black spheres represent Ag(1), Ag(2), Pb, and O ions, respectively.
Two possible cleavage planes with Ag(1) and Ag(2) terminations are shown as a dashed line.
}
\end{figure}

Since the discovery of the high-temperature superconducting cuprate,\cite{firstcuprate} two-dimensional layered oxide superconductors have become a central research subject in condensed matter physics.\cite{cupratereview}
Ag$_5$Pb$_2$O$_6$ was recently added to the list.
Figure~1 shows its layered crystal structure, in which the PbO$_3$ honeycomb layer is sandwiched between two types of Ag layers. 
Despite its low $T_{\rm c}$ of 52.4~mK,\cite{superconductivity} the material has attracted significant attention because of its nearly-free-electron nature in the normal metallic phase.\cite{LDA,dHvA}
Typical nearly-free-electron materials, such as alkali and noble metals, do not exhibit superconductivity at ambient pressure, except Li, which has an extremely low $T_{\rm c}$ of 0.4~mK.\cite{Li} 
Therefore, Ag$_5$Pb$_2$O$_6$ has been considered an ideal model system for investigating superconductivity without the effects of the complicated electronic structure.

For superconductors, understanding which electrons will participate in Cooper pairing is important.
However, the conduction-state formation of Ag$_5$Pb$_2$O$_6$ remains controversial.
Assuming the formal valencies of Ag, Pb, and O ions of +1, +4, and -2, respectively, the material should have one excess electron per chemical formula, which will contribute to free-electron-like conduction.
A simple look at the crystal structure suggests that the excess electron may belong to the Ag layers.
An earlier extended H\"{u}ckel tight-binding study also suggested that only the Ag 5$s$ orbital contributes to the conduction of the material.\cite{EHTB}
In contrast, a recent theoretical study using local density approximation (LDA) suggested a different electronic structure, wherein a Pb 6$s$ orbital hybridized with O 2$p$ orbitals crosses the Fermi level ($E_{\rm F}$), forming a nearly-free-electron band.\cite{LDA}

Several experimental studies have attempted to understand the valence state of this material.
Some bond length analyses were performed but provided contradicting results regarding the distinction between Ag-dominant\cite{oldstructure,oldbond} and Pb-dominant\cite{XAS} characters, while de Haas-van Alphen measurements\cite{dHvA} demonstrated a nearly-free-electron Fermi surface and indirectly supported the validity of the LDA scenario\cite{LDA}.
The only direct experimental evidence was from X-ray absorption spectroscopy (XAS), which suggested that the conduction electrons may be formed by the Pb 6$s$ orbital.\cite{XAS}
However, the absorption edge in the XAS was not sharp enough to provide decisive experimental evidence.

Valence-band photoemission measurements have served as an ideal tool for identifying the orbital character because they can directly probe valence bands of a solid.
Each orbital has a distinct photon-energy dependence in photoionization cross sections.\cite{crosssection}
By varying the incident photon energies, photoemission spectra can be tuned to obtain the orbital-selective partial density of states (DOS) of valence bands.\cite{crosssecexp}
However, in case of Ag$_5$Pb$_2$O$_6$, the photon-energy dependence of photoionization cross sections of Ag 5$s$ and Pb 6$s$ orbitals are quite similar.\cite{crosssection}
Therefore, investigating the orbital characters of Ag$_5$Pb$_2$O$_6$ using valence-band photoemission is difficult.

Here, we present an alternative approach that utilizes core-level photoemission to investigate the element-specific orbital characters of the conduction electrons in Ag$_5$Pb$_2$O$_6$.
We show that the detailed core-level lineshape, especially its asymmetry, is strongly influenced by the Coulomb interaction between a core hole and conduction electrons.
We observe strong asymmetry in a Pb 4$f_{7/2}$ core-level spectrum, while a Ag 3$d_{5/2}$ spectrum is nearly symmetric.
This difference implies that the Pb 6$s$ conduction electrons dominate near $E_{\rm F}$.
We also demonstrate that the asymmetry of the core-level spectrum can be understood quantitatively based on the Mahan-Nozi\`{e}res-DeDominicis (MND) theory\cite{M,ND,MNDReview}.
The Pb 4$f_{7/2}$ spectrum of Ag$_5$Pb$_2$O$_6$ cannot be explained by the Doniach-\v{S}unji\'{c} (D\v{S}) lineshape\cite{DS}, which has been widely used to explain the core-level spectra of many metallic materials. 
By entering a realistic DOS of the Pb 6$s$ orbital\cite{LDA} into the MND model, we can successfully explain the lineshape quantitatively. 
Our fit results support the LDA results; i.e., Ag$_5$Pb$_2$O$_6$ has a free-electron-like electronic structure but with a narrow bandwidth.

\section*{Results and Discussion}

\begin{figure}
\centering
\includegraphics{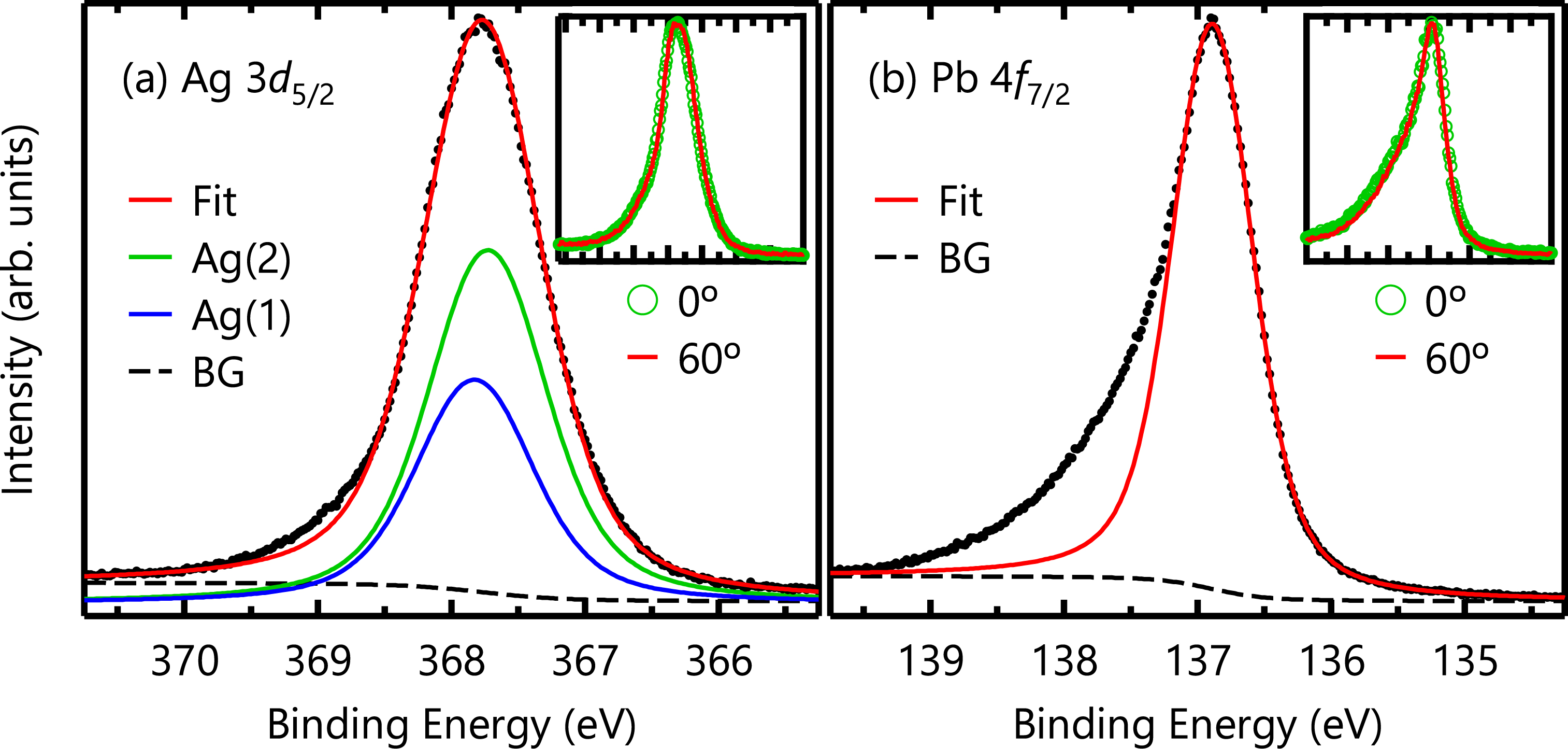}
\caption
{
(a) Ag 3$d_{5/2}$ and (b) Pb 4$f_{7/2}$ core-level spectra of Ag$_5$Pb$_2$O$_6$.
Red lines represent results fitted by symmetric Lorentzian peaks.
The insets show the takeoff-angle dependence in each spectrum.
The negligible differences imply that there are no surface states in either spectra.
}
\end{figure}

The Ag 3$d_{5/2}$ and Pb 4$f_{7/2}$ core-level spectra of Ag$_5$Pb$_2$O$_6$ are shown in Figs.~2(a) and (b), respectively.
Both spectra are fitted by symmetric Lorentzian peaks convoluted by a Gaussian after subtracting Shirley backgrounds\cite{Shirley}.
Red lines in Fig.~2 represent the results of the fitting.
To verify any surface contributions, we measured core-level spectra at takeoff angles of $0^\circ$ and $60^\circ$.
At $60^\circ$, the effective probing depth should be half that of $0^\circ$.
Thus, in the spectrum with a $60^\circ$ takeoff angle, a surface peak should be much more intense than the bulk peaks.
However, as shown in the insets of Fig.~2, the spectral differences between the $0^\circ$ and $60^\circ$ spectra are negligible, which confirms the absence of surface-state contributions.

As shown in Fig.~2(a), Ag 3$d_{5/2}$ core-level spectrum can be well described by two symmetric peaks with a Lorentzian full-width at half-maximum (FWHM) of 0.274~eV\cite{coreref}, which we attribute to two crystallographic Ag sites: Ag(1) and Ag(2), as schematically displayed in Fig.~1. 
The intensity ratio between these two peaks can be determined using the emission depth distribution function\cite{Hufner}. 
We calculated the inelastic mean free path using the TPP-2M equation\cite{TPP} embedded in the NIST Electron Inelastic-Mean-Free-Path Database (https://www.nist.gov/srd/nist-standard-reference-database-71). 
Using the value of 4.93~\AA\ for the inelastic mean free path, we evaluated the emission depth distribution function.
With a calculated ratio of Ag(1) : Ag(2) = 1 : 1.58, we could obtain a good fit, as shown in Fig.~3(a). 
The small discrepancy between the experimental and fit results may be due to the imperfection of the sample surface. 
Conversely, the Pb 4$f_{7/2}$ core-level spectrum cannot be explained by a single symmetric peak, leaving a large amount of spectral weight at high binding energies, as shown in Fig.~2(b). 

\begin{figure}
\centering
\includegraphics{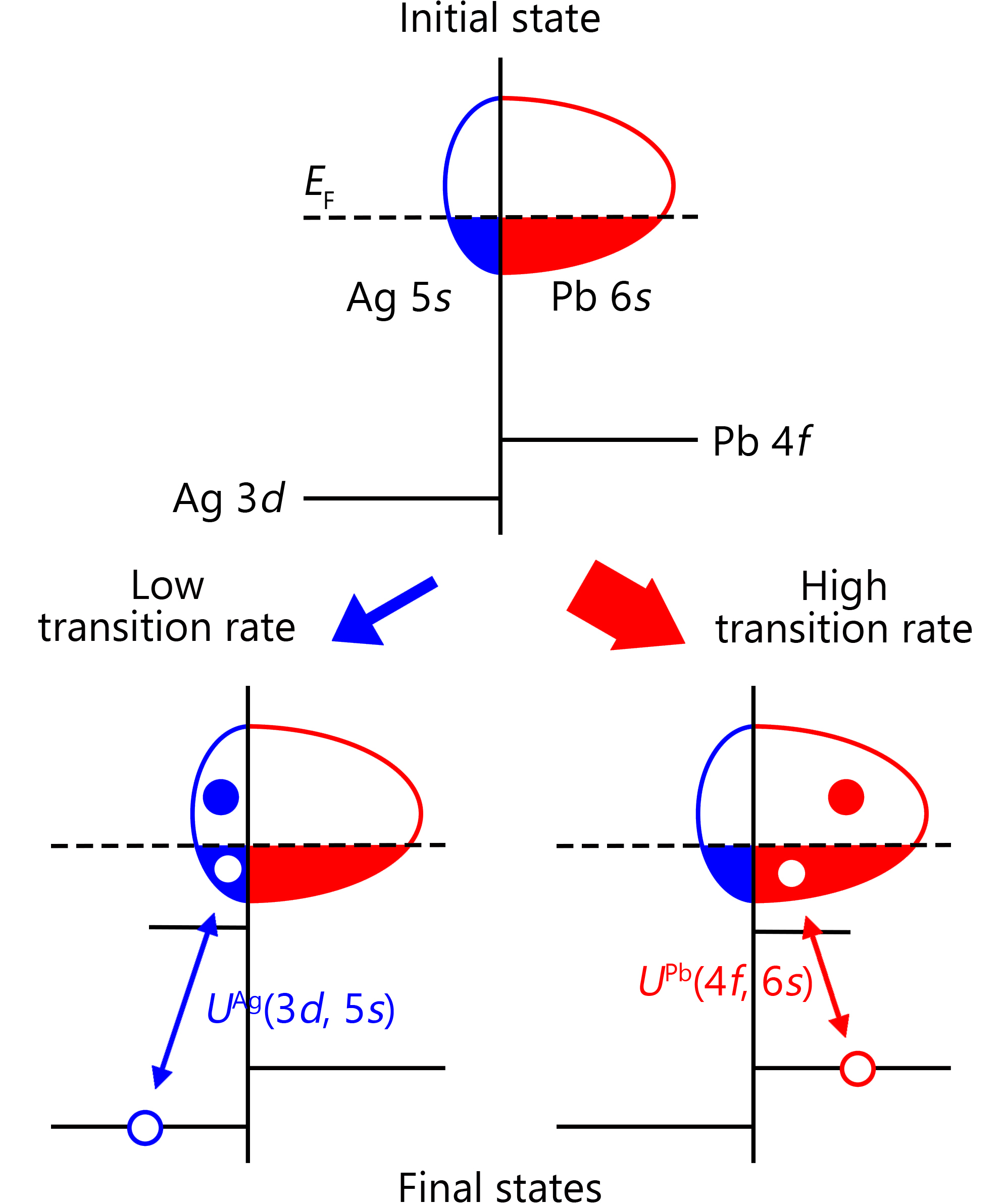}
\caption
{
Schematic diagram of core-level photoemission in Ag$_5$Pb$_2$O$_6$.
A Pb-dominant character is assumed for the conduction electrons in the diagram.
A Coulomb interaction between a Pb core hole and Pb conduction electrons generates electron-hole pairs in the $E_{\rm F}$-crossing band.
}
\end{figure}

The strong asymmetry of the Pb 4$f_{7/2}$ spectrum should be directly related to dominant Pb 6$s$ partial DOS near the $E_{\rm F}$. 
Figure~3 schematically shows the core-level photoemission processes in Ag$_5$Pb$_2$O$_6$. 
Local Coulomb interaction by a core photohole will scatter conduction electrons, thus exciting electron-hole pairs in a $E_{\rm F}$-crossing band. 
Due to the local attractive core-hole potential, an impurity-like state appears below the bottom of the valence band. 
The energies of the final states with electron-hole pair excitations are higher than that of the lowest energy state, resulting in a large spectral weight at a high binding energy and an asymmetric core-level lineshape. 
In Ag$_5$Pb$_2$O$_6$, we observed that a Ag 3$d_{5/2}$ spectrum is nearly symmetric, indicating few excitations at the conduction band.
In contrast, the Pb 4$f_{7/2}$ core-level spectrum is highly asymmetric, which implies that many electron-hole pairs are excited at the Pb 6$s$ conduction band, as shown in Fig.~3.
Therefore, the main character of the conduction electrons in Ag$_5$Pb$_2$O$_6$ should be the Pb 6$s$ electrons.

The asymmetric core-level lineshape in a metal can be understood based on the MND theory,\cite{M,ND,MNDReview} in which a local Coulomb potential due to a core hole scatters conduction electrons.
The model Hamiltonian is given by
\begin{equation}
H = \sum_{\bf k} \epsilon_{\bf k} c_{\bf k}^\dagger c_{\bf k} + \epsilon_h h^\dagger h - \frac{U}{N} \sum_{\bf k k^\prime} c_{\bf k}^\dagger c_{\bf k^\prime} h^\dagger h,
\end{equation}
where $c_{\bf k}^\dagger (c_{\bf k})$ is an conduction-electron creation (annihilation) operator, $h^\dagger (h)$ a core-hole creation (annihilation) operator, $\epsilon_{\bf k}$ a conduction-electron energy with a momentum {\bf k}, $\epsilon_h$ a core-hole energy, and $N$ the number of lattice sites considered in the model.
$U$ is an attractive on-site Coulomb interaction between conduction electrons and a core hole.
Without $U$, the core-level spectrum should have a simple Lorentzian lineshape, whose half width at half-maximum $\gamma$ reflects a core-hole lifetime.
By turning on $U$, electron-hole pairs can be excited across the $E_{\rm F}$, resulting in an asymmetric lineshape.
Thus, a quantitative analysis of an asymmetric spectral function $A(\omega)$ can provide us with important information regarding conduction bands, such as $U$ and DOS near the $E_{\rm F}$.

If one assumes that the conduction-band DOS is flat, and the bandwidth is infinite, $A(\omega)$ can be expressed in a closed-form expression in the MND theory, which yields the well-known D\v{S} lineshape\cite{DS}:
\begin{equation}
A_\textrm{D\v{S}}(\omega) = \frac{\Gamma(1-\alpha)\cos\Big[\frac{\pi\alpha}{2}+(1-\alpha)\tan^{-1}(\frac{\omega}{\gamma})\Big]}{(\omega^2+\gamma^2)^{(1-\alpha)/2}},
\end{equation}
where $\Gamma(x)$ is the gamma function and $\alpha$ is an asymmetry parameter.
When $\alpha=0$, the lineshape is just a symmetric Lorentzian of half-width $\gamma$. 
According to the MND theory, $\alpha$ is approximately given by $U^2\rho^2(E_{\rm F})$ at the weak-coupling limit, where $\rho(\epsilon)$ is the conduction-band DOS.\cite{DS}
Because of the strong locality of $U$, the strength of asymmetry $\alpha$ should be related to an element-specific partial DOS.
Note that the D\v{S} lineshape has been quite successful at describing core-level spectra from numerous metals, such as noble\cite{nobleDS} and simple\cite{simpleDS1,simpleDS2} metals.

\begin{figure}
\centering
\includegraphics{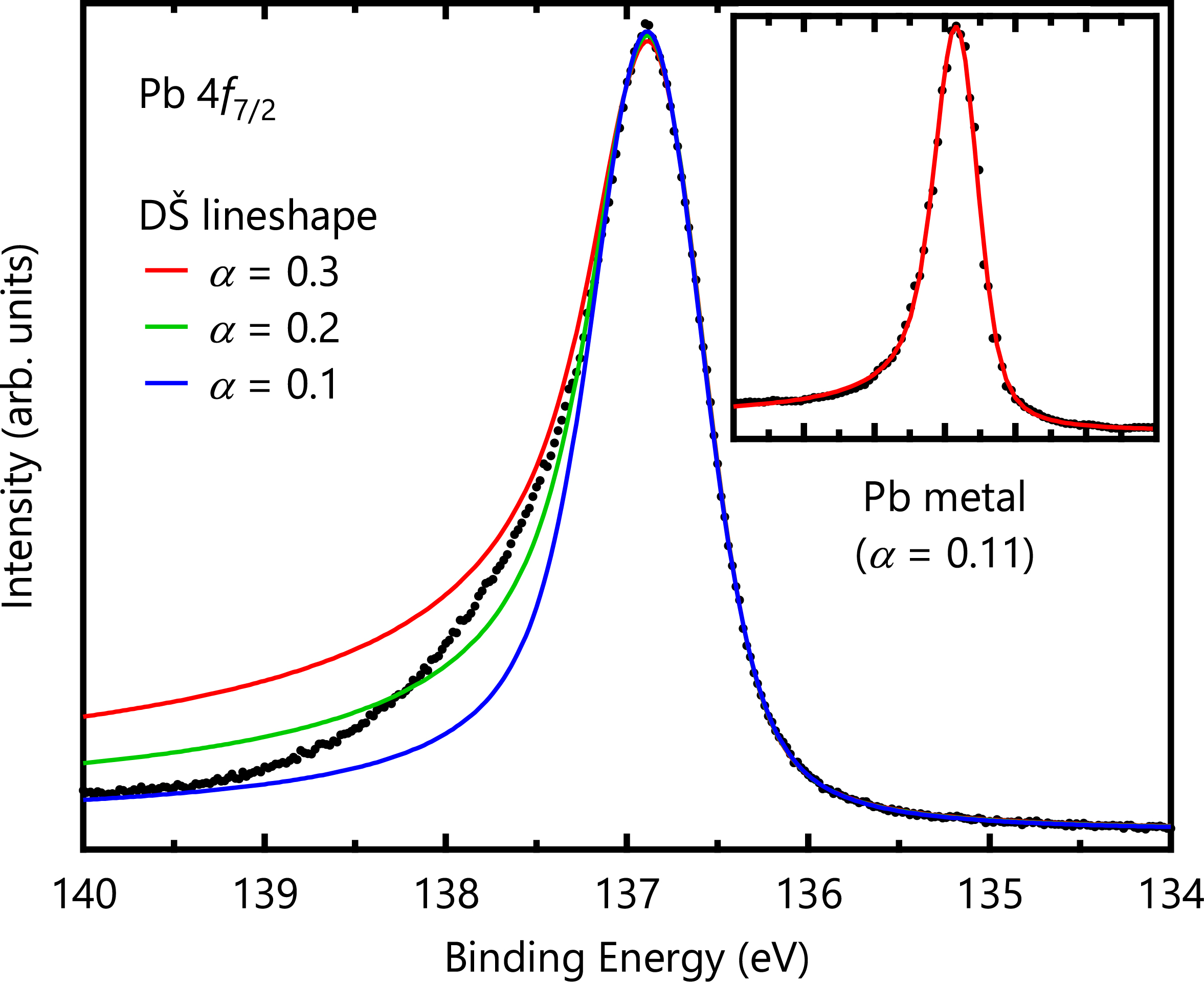}
\caption
{
Pb 4$f_{7/2}$ core-level spectrum and fitting results of the D\v{S} lineshape with several  values.
The inset shows a Pb 4$f_{7/2}$ core-level spectrum of a Pb metal from Ref.~\citenum{Pbref}, which is well explained by the D\v{S} lineshape with $\alpha=0.11$.
}
\end{figure}

However, the measured Pb 4$f_{7/2}$ spectrum is not explained by the simple D\v{S} lineshape.
Fig.~4 shows several trial fits of the Pb 4$f_{7/2}$ core-level spectrum wherein the value of $\alpha$ using Eq. (2) is varied. 
When $\alpha$ is small, the large asymmetry cannot be explained. 
Conversely, large $\alpha$ values render large tails at higher binding energies. 
The inset of Fig.~4 shows a Pb 4$f_{7/2}$ core-level spectrum of the Pb metal from Ref.~\citenum{Pbref}, which is well explained by the D\v{S} lineshape with $\alpha=0.11$, unlike the case of Ag$_5$Pb$_2$O$_6$. 
This discrepancy may be ascribed to the assumption of the flat, broad conduction-band DOS used to obtain the D\v{S} lineshape. 
A similar discrepancy in the D\v{S} lineshape behavior has also been reported in narrow $d$-band metals such as Pd and Pt.\cite{nobleDS} 

To explain the asymmetric lineshape at a quantitative level, we adopted a theoretical approach reported by Davis and Feldkamp\cite{Davis,DavisReview}.
The model reads
\begin{equation}
H = \int\! \epsilon \psi^\dagger_\epsilon \psi_\epsilon\, d\epsilon + \epsilon_h h^\dagger h - \iint\! \tilde{U}(\epsilon,\epsilon^\prime) \psi^\dagger_\epsilon \psi_{\epsilon^\prime} h^\dagger h\, d\epsilon d\epsilon^\prime,
\end{equation}
where $\psi^\dagger_\epsilon (\psi_\epsilon)$ is a creation (annihilation) operator of a conduction electron with an energy $\epsilon$. Here, $\tilde{U}$ is given by
\begin{equation}
|\tilde{U}(\epsilon,\epsilon^\prime)|^2 = U^2 \rho(\epsilon) \rho(\epsilon^\prime).
\end{equation}
After the creation of a core hole, a final one-electron eigenstate $\phi_n$ with an energy eigenvalue $\omega_n$ is given by a unitary transformation $S_n(\epsilon)$ of the initial eigenstate $\psi_\epsilon$: i.e.
\begin{equation}
\phi_n = \int\! S_n(\epsilon) \psi_\epsilon\, d\epsilon.
\end{equation}
After discretizing the energy for numerical calculations, a final many-body state $\Phi_\alpha$ is given by a product of $L$ eigenstates: $\Phi_\alpha = \phi_{n_1} \phi_{n_2} \cdots \phi_{n_L}$, with an energy of $E_\alpha = \omega_{n_1}+\omega_{n_2}+\cdots+\omega_{n_L}$, while its spectral intensity is given by $|\langle\Phi_\alpha|\Phi_G\rangle|^2$, where $\Phi_G$ corresponds to the initial many-body ground state with the lowest occupied $L$ eigenstates.
Note that $|\langle\Phi_\alpha|\Phi_G\rangle|^2$ corresponds to the square of a determinant of an order-$L$ matrix whose elements are $S_n(\epsilon)$.
When $\rho(\epsilon)$ is given, we can easily calculate $S_n(\epsilon)$ and its associated spectral intensity.
Then, the spectral function $A(\omega)$ is given by the total sum of spectral contributions of the $_N{\rm C}_L$ final states. 

\begin{figure}
\centering
\includegraphics{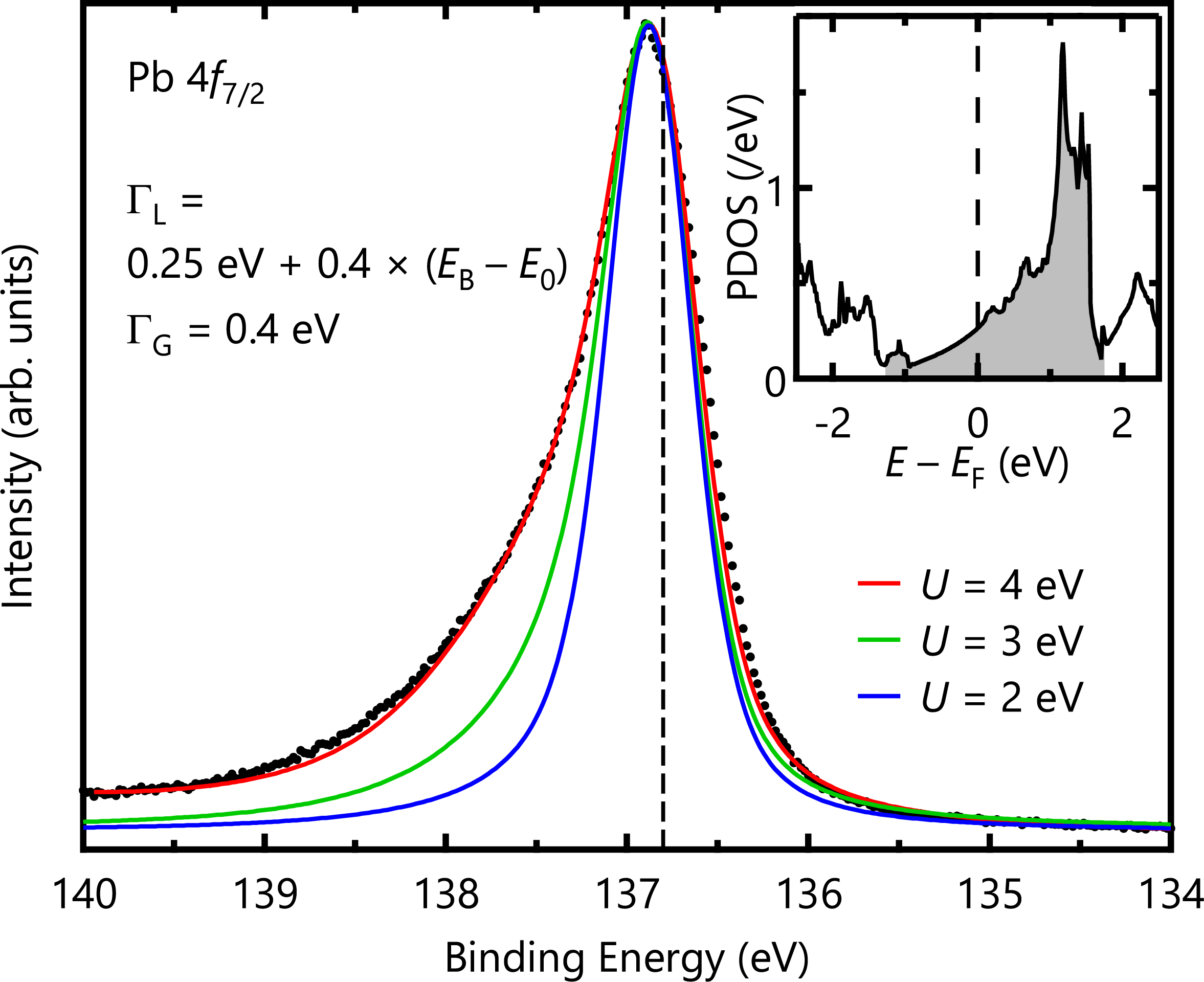}
\caption
{
Pb 4$f_{7/2}$ core-level spectrum and fitting results of MND-model calculations with the Pb 6$s$ partial DOS.
The spectrum is well explained with $U = 4$~eV.
The dashed line represents the lowest binding energy of 136.8~eV.
The inset shows the Pb 6$s$ partial DOS from LDA calculations, and the gray shaded region was considered in our calculations.
}
\end{figure}

To explain the Pb 4$f_{7/2}$ spectrum, we calculated the spectrum by employing the realistic Pb 6$s$ partial DOS from LDA calculations\cite{LDA}.
An experimental DOS from valence-band photoemission would not be useful, because it is difficult to extract the contribution from Ag and Pb from the partial DOS of $E_{\rm F}$-crossing bands by varying the incident phonon energy.
Moreover, the photoemission spectrum does not provide information above $E_{\rm F}$.
Instead, we employed the Pb 6$s$ partial DOS from LDA calculations, as shown in the inset of Fig.~5.
According to the MND theory, completely filled or empty conduction bands are not responsible for the asymmetric lineshape, because there is a finite energy gap for electron-hole pair excitations, and their spectral contributions are negligible compared with those from excitations in $E_{\rm F}$-crossing bands.
To calculate a Pb 4$f_{7/2}$ core-level spectrum, we used the partial DOS from -1.27 to 1.72~eV (the gray shaded region in the inset of Fig. 5), which corresponds to the minimum or maximum of the $E_{\rm F}$-crossing band, respectively.\cite{LDA}
In our calculations, final states with one or two electron-hole pair creations are considered to have a total spectral weight larger than 0.999.

The observed Pb 4$f_{7/2}$ spectrum is quantitatively well explained by the MND calculations, as shown in Fig.~5.
To compare it with the experimental spectrum, we convoluted them using a Lorentzian and a Gaussian. 
For the Lorentzian broadening, we used $\Gamma_{\rm L} = 0.25~{\rm eV} + 0.4 \times\ (E_{\rm B} - E_0)$ (FWHM), where the first (second) term corresponds to a lifetime for the core hole (electron-hole pair).
$E_0$ is the lowest binding energy in the calculations. 
The $E_0$ value was set to 136.8~eV for the Pb 4$f_{7/2}$ spectrum, as represented by the dashed line in Fig.~5. 
To account for the extrinsic broadening, we also included a Gaussian broadening with $\Gamma_{\rm G} = 0.4$~eV (FWHM) in the calculations. 
The calculated spectrum with $U = 4$~eV reproduced the observed Pb 4$f_{7/2}$ spectrum quite well. 
The lineshape of the high-binding-energy tail is strongly governed by the DOS shape itself.\cite{DOSShape} 
Therefore, the success of our MND calculations supports the validity of the LDA calculations\cite{LDA}, which suggested the Pb 6$s$ character of charge carriers and the free-electron-like electronic structure.

For comparison, we also depict calculation results with $U = 2$ and 3~eV in Fig.~5.
The figure clearly shows that the asymmetries of the calculated spectra are quite sensitive to the values of $U$.
Since the asymmetry parameter is approximately given by $U^2\rho^2(E_{\rm F})$ in the D\v{S} lineshape,\cite{DS} asymmetry in a core-level spectrum is a good indicator of an element-specific contribution to metallic behavior at similar values of $U$.
According to Hartree-Fock calculations for atomic orbitals using Cowan’s code\cite{Cowan}, the average values of bare Coulomb interaction $U_0$ between a core hole and conduction electrons for Ag 3$d$-5$s$ and Pb 4$f$-6$s$ electrons are 
\begin{equation}
U_0^{\rm Ag}(3d,5s) = F^0_{\rm Ag}(3d,5s)-1/10~G^2_{\rm Ag}(3d,5s) = 11.7~{\rm eV},
\end{equation}
and
\begin{equation}
U_0^{\rm Pb}(4f,6s) = F^0_{\rm Pb}(4f,6s)-1/14~G^3_{\rm Pb}(4f,6s) = 14.9~{\rm eV},
\end{equation}
respectively, where $F^0$, $G^2$, and $G^3$ are Slater integrals\cite{Slater}.
These values of the bare interactions are considerably reduced due to the screening by conduction electrons.
Using the relation between the screened Coulomb interaction $U$ and a polarization $P$, $U=U_0/(1-PU_0)$,\cite{Fetter} we can estimate the value of $U^{\rm Ag}(3d,5s)$ to be 3.7~eV from $U^{\rm Pb}(4f,6s) = 4$~eV, assuming the same polarization for each core hole.
Those values are so similar that we can expect a highly asymmetric lineshape for the Ag 3$d_{5/2}$ spectrum, in contrast to the real one.
Thus, we can conclude that the Ag 5$s$ partial DOS at $E_{\rm F}$ should be much smaller than that of Pb 6$s$.

\section*{Conclusion}

We settled the controversy surrounding the element-specific orbital character of Ag$_5$Pb$_2$O$_6$ by measuring its core-level photoemission spectra.
The dominant Pb 6$s$ character was confirmed based on the strong asymmetry of a Pb 4$f$ core-level spectrum and MND-model calculations.
Our approach using core-level photoemission can be applied to investigate the conduction state of other compound materials, where valence-band photoemission is difficult to apply.
Moreover, the failure of the D\v{S} lineshape indicates that in general, we should use the more realistic DOS to explain the core-level lineshape of materials. 

\section*{Methods}

We synthesized Ag$_5$Pb$_2$O$_6$ single crystals using a previously reported self-flux method\cite{firstsingle}.
First, we mixed AgNO$_3$ and Pb(NO$_3$)$_2$ powders thoroughly and placed the mixture into an alumina crucible.
We initially heated the mixture in air to 90~$^\circ$C and maintained that temperature for 3~hours.
Then, we heated it at 405~$^\circ$C for 4~days.
We varied the ratio between AgNO$_3$ and Pb(NO$_3$)$_2$ to obtain large crystals with a hexagonal stick shape.
We used the X-ray diffraction technique to confirm the crystal structure of the synthesized crystal.
The diffraction pattern indicated that there was no secondary phase.
We also confirmed the stoichiometry between Ag and Pb ions via energy dispersive spectroscopy.

We performed core-level photoemission experiments at the Beamline 4A1 of Pohang Light Source.
The single crystal was cleaved and measured at 50~K under a vacuum of $3~\times~10^{-11}$~Torr.
Possible cleavage planes are the Ag(1) and Ag(2) terminations, as displayed in Fig.~1(a).
To achieve the best energy resolution, we used photon energies of 480~eV for the Ag 3$d_{5/2}$ level and 280~eV for the Pb 4$f_{7/2}$ level.
We determined the binding energy with reference to the 4$f_{7/2}$ core-level peak of polycrystalline gold electrically connected to the sample.

\bibliography{Ref}

\section*{Acknowledgements}

We thank T. Oguchi for providing the electronic-structure data of Ag$_5$Pb$_2$O$_6$.
We also acknowledge the experimental support provided by H. K. Yoo and C.-T. Kuo.
This work was supported by IBS-R009-D1, IBS-R009-G2, and grants from the National Research Foundation of Korea (NRF) (2012M2B2A4029470, 2014R1A1A1002868).
Experiments at PLS-II were supported in part by MSIP and POSTECH.

\section*{Author contributions statement}

S.S. and H.-D.K. conceived and designed the work.
K.D.L., C.J.W., and N.J.H synthesized Ag$_5$Pb$_2$O$_6$ single crystals.
S.S., J.S.O., H.-D.K., and B.-G.P. carried out the core-level photoemission experiments.
Fitting and analyses were conducted by S.S. and H.-D.K.
The research was carried out with guidance from M.H., Y.J.C., C.K., H.-D.K., and T.W.N.
S.S., H.-D.K., and T.W.N. co-wrote the manuscript.
All authors discussed the work and reviewed the manuscript.

\section*{Additional information}

\textbf{Competing financial interests}: The authors declare no competing financial interests.

\end{document}